\documentclass[11pt]{article}
\usepackage{graphicx} 
\usepackage{amssymb,amsthm}  
\usepackage{cite}   

\begin{document}

\def \d {{\rm d}}
\def \U {{\cal U}}
\def \V {{\cal V}}
\def \H {{\cal H}}
\def \M {{\cal M}}
\def \e {\varepsilon}
\def \E {{\bf e}}
\def \m  {{\bf m}}
\def \bl {{\bf l}}
\def \k  {{\bf k}}
\def \uu {{\bf u}}
\def \R  {{\cal R}}
\def \I  {{\cal I}}
\def \A  {{\cal A}}
\def \C  {{\cal C}}
\newcommand{\be}{\begin{equation}}
\newcommand{\ee}{\end{equation}}
\newcommand{\beqn}{\begin{eqnarray}}
\newcommand{\eeqn}{\end{eqnarray}}
\newcommand{\pa}{\partial}

\newcommand{\pp}{{\it pp\,}-}


\def \E {{\bf e}}
\def \m  {{\bf m}}
\def \mc  {{\bf\bar m}}
\def \bl {{\bf l}}
\def \k  {{\bf k}}
\def \uu {{\bf u}}

\def \a {\alpha}
\def \ac {\bar \alpha}
\def \b {\beta}
\def \bc {\bar \beta}
\def \g {\gamma}
\def \gc {\bar \gamma}
\def \e {\varepsilon}
\def \ec {\bar \varepsilon}
\def \kk {\kappa}
\def \kkc {\bar \kappa}
\def \l {\lambda}
\def \lc {\bar \lambda}
\def \mm {\mu}
\def \mmc {\bar \mu}
\def \n {\nu}
\def \nc {\bar \nu}
\def \p {\pi}
\def \pc {\bar \pi}
\def \rr {\rho}
\def \rc {\bar \rho}
\def \s {\sigma}
\def \sc {\bar \sigma}
\def \t {\tau}
\def \tc  {\bar \tau}

\def \dec {\bar \delta}
\def \de {\delta}
\def \D {\Delta}

\def \phoo {\Phi_{11}}
\def \pho {\Phi_1}
\def \phoc {\bar\Phi_1}
\def \phtt {\Phi_{22}}
\def \pht {\Phi_2}
\def \phtc {\bar\Phi_2}
\def \pn {\Psi_0}
\def \po {\Psi_1}
\def \pt {\Psi_2}
\def \ptt {\Psi_3}
\def \pf {\Psi_4}

\newtheorem*{theorem}{Theorem}

\title{Explicit Kundt type~$II$ and $N$ solutions as gravitational waves in various
   type~$D$ and $O$ universes}

\addvspace{1cm}

\author{
J. Podolsk\'y\thanks{E--mail: {\tt podolsky@mbox.troja.mff.cuni.cz}}
\\ Institute of Theoretical Physics, Charles University in Prague,\\
V Hole\v{s}ovi\v{c}k\'ach 2, 18000 Prague 8, Czech Republic.
 \\ \\
and \\ \\
M. Ortaggio\thanks{E--mail: {\tt ortaggio@science.unitn.it}} \\
Dipartimento di Fisica, Universit\`a degli Studi di Trento,  \\
and INFN, Gruppo Collegato di Trento, 38050 Povo (Trento), Italy. \\ \\
}

\date{\today}

\maketitle
\begin{abstract}
A particular yet large class of non-diverging solutions which admits a cosmological constant,
electromagnetic field, pure radiation and/or general non-null matter component is
explicitly presented. These spacetimes represent exact gravitational waves
of arbitrary profiles which propagate in  background universes
such as Minkowski, conformally flat (anti-)de Sitter, Edgar--Ludwig, Bertotti--Robinson,
and type $D$ (anti-)Nariai or Pleba\'nski--Hacyan spaces, and their
generalizations. All  possibilities are discussed
and are interpreted  using a unifying simple metric form.
Sandwich and impulsive waves propagating in the above background
spaces with different geometries and matter content can easily be constructed.
New solutions are identified, e.g. type $D$ pure radiation or
explicit type $II$ electrovacuum waves in (anti-)Nariai universe.
It is also shown that, in general, there are no conformally flat
Einstein--Maxwell fields with a non-vanishing cosmological
constant.


\end{abstract}

\newpage

\section{Introduction}

The Kundt class, which is characterized by the geometrical property that it admits a non-expanding
and non-twisting null congruence, is a well-known family of solutions to Einstein's equations.
Wolfgang Kundt was the first to introduce, emphasize and investigate this large class
\cite{kundt1, kundt62, EK} (in the case of vacuum and pure radiation)
although some of its important subclasses, in particular the {\it pp\,}-waves
\cite{brinkmann,BalJef26,Brdicka51,BPR} or the Nariai \cite{Nariai51} and Bertotti-Robinson
\cite{LeviCivita17BR,Bertotti59,Robinson59} universe, were discovered and studied previously.
Since then a great number of papers have been devoted to the derivation of such spacetimes and
an analysis of their properties, see e.g. \cite{kramerbook} for the review. More
recent articles which are  related to the specific topic of this paper are mentioned below
 in the appropriate context.

Naturally, most  works on exact non-diverging spacetimes investigate particular subclasses
of the large Kundt family by restricting attention to a specific algebraic Petrov type {\em and} considering
special matter contents (vacuum, cosmological constant, pure radiation, null or non-null
electromagnetic fields). In many of these subcases, all solutions of the given type were explicitly
obtained. On the other hand, at present it is still impossible to find a general solution which
describes different Petrov types and matter fields in an explicit form. Even
with an assumption of symmetries such solutions could in most cases
only be given implicitly, see e.g. \cite{kramerbook,MacSik92}.

This contribution focuses on the ``gap'' which lies between these
``extreme'' approaches.  We present in section~2 a  simple explicit  metric which contains
several arbitrary parameters and functions. For particular choices we recover, in sections~3 and 4
respectively, many  Kundt spacetimes with various matter contents
which are of the Petrov types $O$, $D$, and $N$, $II$. Moreover, the latter solutions can be understood
as exact gravitational waves of arbitary profile which propagate in the corresponding
``background'' universes. All  possibilities are discussed below in
detail  and are interpreted by analyzing the geodesic deviation. The paper
also contains an appendix in which we prove that there are no
conformally flat Einstein--Maxwell fields with $\Lambda\not=0$.

\section{General form of the solutions}
Throughout the present paper we consider and discuss the following metric
\be
 \d s^2=2\,{1\over P^2}\,\d\zeta\,\d\bar\zeta-2\,{Q^2\over P^2}\,\d u\,\d v+F\,\d  u^2\ ,
 \label{metric}
\ee
with
\beqn
 P & = & 1+\alpha\,\zeta\bar\zeta\ , \nonumber \\
 Q & = & (1+\beta\,\zeta\bar\zeta)\,\e+C\,\zeta+\bar{C}\,\bar\zeta\ ,  \label{coeff} \\
 F & = & D\,{Q^2\over P^2}\,v^2-\frac{(Q^2)_{,u}}{P^2}\,v-\frac{Q}{P}\,H\ , \nonumber
\eeqn
where $\alpha$, $\beta$, and $\e$ are real constants (without loss of generality we can assume
$\e=0$ or $\e=1$), $C(u)$ and $D(u)$ are arbitrary functions of the null coordinate $u$
($C$ may be complex), and $H(\zeta,\bar\zeta,u)$ is an arbitrary function of the spatial
coordinates $\zeta$, $\bar\zeta$, and of $\,u$. In the natural null tetrad
\beqn
 {\bf m} & = & P\,\pa_{\bar{\zeta}}\ , \nonumber \\
 {\bf\bar m} & = & P\,\pa_{\zeta}\ , \nonumber \\
 {\bf l} & = & {P^4 \over 2Q^4}F\,\pa_v+{P^2\over Q^2}\,\pa_u\ , \label{privtetrad}\\
 {\bf k} & = & \pa_v\ , \nonumber
\eeqn
we obtain (using $\e^2\equiv\e$) the following  form of the only non-vanishing Weyl and Ricci
scalars
\beqn
 \Psi_2 & = & -{\textstyle\frac{1}{6}}\left(D+2\e\beta -2C\bar{C}\right)\frac{P^2}{Q^2}\ , \label{psi2} \\
 \Psi_4 & = & {\textstyle\frac{1}{2}}\,(PH)_{,\zeta\zeta}\,\frac{P^4}{Q^3}\ , \label{psi4} \\
  R & = & 24\alpha-12\e(\alpha+\beta)\,\frac{P}{Q}+2\left(D+2\e\beta -2C\bar{C}\right)\frac{P^2}{Q^2}\ , \label{R}\\
 \Phi_{11}  & = & {\textstyle\frac{1}{2}\e}(\alpha+\beta)\,\frac{P}{Q}-{\textstyle\frac{1}{4}}\left(D+2\e\beta -2C\bar{C}\right)\frac{P^2}{Q^2}\ , \label{phi11} \\
 \Phi_{22}  & = & {\textstyle\frac{1}{2}}\bigg\{\Big[\,\e(\alpha-\beta)\,(1-\alpha\,\zeta\bar\zeta)\,+2\alpha\,(C\,\zeta+\bar{C}\,\bar\zeta)\,\Big]\,H+P^2Q\,H_{,\zeta\bar\zeta} \nonumber \\
    & & \hskip 20mm{}-2\e(\alpha+\beta)\,v\,\,(\dot{{C}}\,\zeta+\dot{\bar{C}}\,\bar\zeta)\bigg\}\,\frac{P^3}{Q^4}\ . \label{phi22}
\eeqn
The spin coefficients $\rho$, $\sigma$ and $\kappa$ vanish identically so that the multiple principal null
direction ${\bf k}$ is non-expanding, twist-free, shear-free and geodesic. The spacetimes  (\ref{metric}),
(\ref{coeff}) thus belong to the Kundt class and are algebraically special.
By a simple transformation $v=(P^2/Q^2)\,w$, these can be put into the Kundt canonical form \cite{kramerbook}
 \be
 \d s^2=2\,{1\over P^2}\,\d\zeta\,\d\bar\zeta-2\,\d u\,\left(\d w+W\,\d\zeta+\bar{W}\d\bar\zeta+{\cal H}\,\d u\right)\ ,
 \label{metricK}
\ee
where
\beqn
 W & = & 2w\,\frac{QP_{,\zeta}-PQ_{,\zeta}}{PQ}\ \equiv\ -2w\,\frac{C-\e(\alpha-\beta)\bar\zeta-\alpha\bar{C}\bar\zeta^2}{PQ}\ , \nonumber\\
 {\cal H} & = & -{\textstyle\frac{1}{2}}D\,{P^2\over
 Q^2}\,w^2-\frac{Q_{,u}}{Q}\,w+\frac{Q}{2P}\,H\ .   \label{coeffK}
\eeqn

It follows from (\ref{psi2}) and (\ref{psi4}) that, in general, the above spacetimes are of Petrov type $II$. In special cases,
these degenerate to types $D$, $N$, or can be conformally flat.
It is convenient to introduce the following two functions,
\beqn
 D_0 & = & -2\e\beta +2C\bar{C}\ , \\
 H_0 & = &
 \frac{A_0+A_1\zeta+\bar{A}_1\bar\zeta+A_2\zeta\bar\zeta}{1+\alpha\,\zeta\bar\zeta}\ ,\label{H0}
\eeqn
where $A_0$ and $A_2$ are arbitrary real functions and $A_1$ is an arbitrary complex function of $u$.
When the functions $D$ and $H$ in (\ref{coeff}) take these special forms $D_0$ and $H_0$, the scalars $\Psi_2$ and $\Psi_4$ vanish,
respectively. We may thus easily summarize the possible  types in the following
table~\ref{petrov}.
\begin{table}[htbp]
 \[
  \begin{array}{|c||c|c|} \hline
    & D=D_0 & D\neq D_0 \\
    & (\Psi_2=0) & (\Psi_2\neq0) \\ \hline\hline
    & & \\
    H=H_0 & O & D \\
   (\Psi_4=0) & & \\ \hline
    & & \\
   H\neq H_0 & N &II \\
   (\Psi_4\neq0)& & \\ \hline
  \end{array}
 \]
 \caption{Possible Petrov types of the spacetimes (\ref{metric}).}
 \label{petrov}
\end{table}

In general, the above spacetimes do not have a constant Ricci curvature $R$. They also contain non-uniform
pure radiation described by $\Phi_{22}$ plus a non-null matter component $\Phi_{11}$ (which may be
associated to a fluid with anisotropic pressure, see \cite{CarZak99},
with the 4-velocity of the fluid ${\bf u}=\frac{1}{\sqrt{2}}({\bf k}+{\bf l})$, and
${\bf n}=\frac{1}{\sqrt{2}}({\bf k}-{\bf l})$ being the vector responsible for the anisotropy),
cf. (\ref{R})--(\ref{phi22}). In particular cases, a constant scalar $R$ may correspond to the cosmological
constant $\Lambda$, and the matter components $\Phi_{11}$, $\Phi_{22}$ could represent an electromagnetic field.
In the latter case, the Maxwell equations (see, e.g. \cite{kramerbook})  for the metric (\ref{metric}), (\ref{coeff}) reduce to
\be
 D\Phi_1 =  0\ , \qquad  D\Phi_2-\bar\delta\Phi_1 = \,2\pi\,\Phi_1\ ,\qquad
 \delta\Phi_1=2\tau\,\Phi_1\ ,\qquad\delta\Phi_2-\Delta\Phi_1=\,(\tau-2\beta)\,\Phi_2\ ,
\label{Maxwell}
\ee
since $\>\Phi_0=0\>$ because $\>\Phi_{ij} \equiv \Phi_i\bar\Phi_j=0\>$ for $i=0$,
$j=0,1,2$. The non-vanishing spin coefficients we need here are given by
$\tau=-\bar\pi=Q\left(P/Q\right)_{,\bar\zeta}$ and
$\tau-2\beta=\left(Q^2/P^2\right)\left(P^3/Q^2\right)_{,\bar\zeta}$.
Moreover, $\>\Phi_{12}=0\>$ so that either $\>\Phi_1\>$ or $\>\Phi_2\>$ must vanish.
There are thus \emph{only two} decoupled possibilities, and the
equations (\ref{Maxwell}) can explicitly be integrated as
\beqn
Case\  1: \qquad &&\Phi_1(\zeta,\bar\zeta)=\,E_1\frac{P^2}{Q^2}\ ,\qquad \Phi_2=\,0\ , \label{E1}\\
Case\  2: \qquad &&\Phi_1=\,0\ ,\hskip25mm \Phi_2(\zeta,\bar\zeta,u)=\,E_2\frac{P^3}{Q^2}\ ,\label{E2}
\eeqn
where $E_1(u)$ and $E_2(u,\zeta)$ are complex functions. However, for the purely electromagnetic case
there are additional constraints since we have to satisfy the
Einstein--Maxwell system. This requires that the Ricci scalar $R$ given by (\ref{R})
is equal to $4\Lambda$, and the expressions for $\Phi_{11}$ and $\Phi_{22}$
given by (\ref{phi11}), (\ref{phi22}) have to satisfy
\be
\Phi_{11}= |E_1|^2\frac{P^4}{Q^4}\ , \qquad
\Phi_{22}= |E_2|^2\frac{P^6}{Q^4}\ , \label{phis}
\ee
respectively. In fact, it can be seen that these restrictions rule out the presence of a
purely electromagnetic field in general spacetimes of the form (\ref{metric}),
(\ref{coeff}). Only  for some special choices of $\alpha$, $\beta$, $\e$,  $C$, $D$ and $H$ are
the Einstein--Maxwell field equations  satisfied.

In the next sections we shall describe all the particular  possibilities in  detail.
We start with various backgrounds which may be either conformally flat or of the type $D$.
Then we shall introduce gravitational waves into all these
background spacetimes by considering the corresponding type $N$ or $II$ solutions.

\section{The possible backgrounds}

\label{sec_backgrounds}

\subsection{Conformally flat spacetimes}

\label{sub_confflat}

The above solutions (\ref{metric}), (\ref{coeff}) are conformally flat if and only if $D=D_0$ and $H=H_0$, in which case $\Psi_2=0=\Psi_4$.
For this choice, the expressions (\ref{R})--(\ref{phi22}) reduce to
\beqn
  R & = & 24\alpha-12\e(\alpha+\beta)\,\frac{P}{Q}\ , \label{Rcon}\\
 \Phi_{11}  & = & {\textstyle\frac{1}{2}\e}(\alpha+\beta)\,\frac{P}{Q}\ \equiv\
      \alpha-{\textstyle\frac{1}{24}}R\ , \label{phi11c} \\
  \Phi_{22}  & = & {\textstyle\frac{1}{2}}\bigg\{(\alpha A_0+A_2)\,Q-\e(\alpha+\beta)(A_0+A_1\zeta+\bar{A}_1\bar\zeta+A_2\zeta\bar\zeta)\, \nonumber \\
    & & \hskip 20mm{}-2\e(\alpha+\beta)\,v\,\,(\dot{{C}}\,\zeta+\dot{\bar{C}}\,\bar\zeta)\bigg\}\,\frac{P^3}{Q^4}
    \ . \label{Phi22con}
\eeqn

In order to obtain only a {\em pure radiation} field, one has to set $\>\beta=-\alpha\>$
so that $\>R=24\alpha\>$, $\alpha={\textstyle\frac{1}{6}}\Lambda$, and
\be
 \Phi_{22} = {\textstyle\frac{1}{2}}(\alpha A_0+A_2)\,\frac{P^3}{Q^3}\ . \label{phi22pure}
\ee
Consequently, we obtain explicit solutions in the form (\ref{metric}) in which
\beqn
 P & = & 1+{\textstyle\frac{1}{6}}\Lambda\,\zeta\bar\zeta\ ,\qquad
 Q  =  (1-{\textstyle\frac{1}{6}}\Lambda\,\zeta\bar\zeta)\,\e+C\,\zeta+\bar{C}\,\bar\zeta\ , \label{coeff1} \\
 F & = & \left({\textstyle\frac{1}{3}}\Lambda \e+2C\bar{C} \right)\,{Q^2\over P^2}\,v^2-\frac{(Q^2)_{,u}}{P^2}\,v-\frac{Q}{P^2}\left(A_0+A_1\zeta+\bar{A}_1\bar\zeta+A_2\zeta\bar\zeta\right)\ , \nonumber
\eeqn
where $A_i$ and $C$ are arbitrary functions of $u$. (Note that $\Phi_{11}$ in the expression (\ref{phi11c})
vanishes also for $\e=0$, which however gives just a subcase of (\ref{coeff1}).)
This complete family of pure radiation, conformally flat spacetimes with $\Lambda$ was mentioned already in \cite{OzsRobRoz85}. As discussed in \cite{OzsRobRoz85,BicPod99I} there exist several invariant
subclasses of this, namely (conformally flat) {\it pp\,} and Kundt waves with $\Lambda=0$, and
their generalizations to non-vanishing $\Lambda$.
For {\it pp\,}-wave spacetimes ($\Lambda=0$, $\varepsilon=1$, $C=0$  so that $P=1=Q$)  or the solution of the Siklos type \cite{Siklos85,Podolsky98sik} ($\Lambda<0$, $\varepsilon=1$, $C=\sqrt{-{1\over6}\Lambda}$), one obtains
$D_0=0$ which implies that $\partial_v$ is a Killing vector.
In particular, this includes the only conformally flat pure radiation  solution of the Einstein--Maxwell equations
with  $\Phi_2(u)=\sqrt{\frac{1}{2}A_2}\,e^{i \varphi(u)}$ in
(\ref{E2}), namely the plane waves introduced by Baldwin, Jeffery \cite{BalJef26}
and Brdi\v cka \cite{Brdicka51} (see the appendix).
Another special  subclass of (\ref{coeff}) arises  for $\e=0$ and $C=1$, in which case $Q=\zeta+\bar\zeta$.
In the canonical coordinates (\ref{metricK}) this corresponds to the metric coefficients~(\ref{coeffK})
\beqn
 W & = & -\frac{2w\left(1-\frac{1}{6}\Lambda\,\bar\zeta^2\right)}{\left(\zeta+\bar\zeta\right)
\left(1+\frac{1}{6}\Lambda\,\zeta\bar\zeta\right)}\ , \label{ELgen}\\
 {\cal H} & = & {\textstyle\frac{1}{2}}\left(A_0+A_1\zeta+\bar{A}_1\bar\zeta+A_2\zeta\bar\zeta\right)\frac{\zeta+\bar\zeta}
{\left(1+\frac{1}{6}\Lambda\,\zeta\bar\zeta\right)^2}-\left(\frac{1+\frac{1}{6}\Lambda\,\zeta\bar\zeta}
{\zeta+\bar\zeta}\right)^2w^2 \ .\nonumber
\eeqn
For vanishing $\Lambda$, this exactly reduces to the interesting class of conformally flat pure radiation metrics
found by Edgar and Ludwig \cite{EdgLud97l,EdgLud97} and discussed e.g. in \cite{Skea97,GriPod98,Barnes01}.

The complementary situation in which there is no pure radiation ($\Phi_{22}=0$) but {\em only a non-vanishing
$\Phi_{11}$ component} of the matter field, requires $\alpha+\beta\neq0$, $\e=1$, $C\equiv\gamma=\,$const.,
$D=D_0=-2\beta+2\gamma\bar \gamma$, $A_2=\beta A_0$
and $A_1=\gamma A_0$, so that $H=A_0 Q/P$. However, this function $H(u)$ can always be
removed by the coordinate transformation $v\to Bv+\dot{B}/D$, $u\to\int B^{-1} du$, for a suitable
choice of $B(u)$. Without loss of generality we may thus write the
solutions of this subclass as the metric (\ref{metric}) in which
\be
 P  =  1+\alpha\,\zeta\bar\zeta\ ,\qquad
 Q  =  1+\beta\,\zeta\bar\zeta+\gamma\,\zeta+\bar{\gamma}\,\bar\zeta\  ,\qquad
 F  =  \left(2\gamma\bar{\gamma}-2\beta \right)\,{Q^2\over P^2}\,v^2\ .  \label{coeff2}
\ee
By performing a simple transformation
$v=r/[\,1+(\beta-\gamma\bar\gamma)ur\,]$,
we can put this into the form
\be
 \d s^2={2\,\d\zeta\,\d\bar\zeta\over P^2}-{Q^2\over P^2}\,{2\,\d u\,\d r\over
 [\,1+(\beta-\gamma\bar\gamma)\,ur\,]^2}\ ,
 \label{metricBR}
\ee
in which $P$ and $Q$ are given by (\ref{coeff2}).
(There is still a coordinate freedom given by
$\zeta\to(\bar b+a\zeta)/(\bar a-\alpha b \zeta)$, $a$ and $b$ being constants, and rescaling of
$u$ and/or $r$, which can be used to modify $\beta$ and $\gamma$.)
 The matter content  in
these spacetimes is described by $\Phi_{11}=\frac{1}{2}(\alpha+\beta)P/Q$ which, in general,
is a function of the spatial coordinate $\zeta$.
However, it immediately follows from (\ref{phi11c}) that the (invariant) {\it additional condition}
$R=\,$const. necessarily implies $R=0$, $\alpha=\beta$, $\gamma=0$, $P=Q$.  Thus, for
a constant $R$ we  obtain the Bertotti--Robinson universe \cite{LeviCivita17BR,Bertotti59,Robinson59,kramerbook}
\be
 \d s^2={2\,\d\zeta\,\d\bar\zeta\over \left(1+\alpha\,\zeta\bar\zeta\right)^2}-2\,\d u\,\d v-2\alpha\, v^2\,\d u^2 \ .
 \label{br}
\ee
This  homogeneous space is the unique conformally flat solution of the Einstein--Maxwell equations
with a (uniform) non-null electromagnetic field
$\Phi_1=\sqrt{\alpha}\,e^{i \varphi}$ in (\ref{E1}) where $\alpha>0$ (see the appendix).

Of course, conformally flat spacetimes of the above form exist which combine $\Phi_{11}$ and  $\Phi_{22}$. For example, one can introduce a pure radiation  in the Bertotti--Robinson universe by adding the term $-H_0\,\d u^2$ to the metric (\ref{br}).

Finally, the conformally flat spaces with {\em no matter} (Einstein spaces) are given by (\ref{coeff1}) with the additional
constraint $A_2=-\frac{1}{6}\Lambda A_0$, see (\ref{phi22pure}). In such a case, the function $H_0$ takes the form
$H_0=[(1-\frac{1}{6}\Lambda\,\zeta\bar\zeta)A_0+A_1\zeta+\bar{A}_1\bar\zeta]/P$, which can be removed completely by
a suitable coordinate transformation \cite{BicPod99I}. These constant curvature vacuum solutions with $H=0$
can then be put into the standard form of Minkowski, de Sitter or anti-de Sitter spacetimes \cite{PodolskyPhD}.
For example, assuming $\e=1$, $C=0$, we obtain
\be
 \d s^2={2\,\d\zeta\,\d\bar\zeta\over \left(1+\frac{1}{6}\Lambda\,\zeta\bar\zeta\right)^2}
 -2\left(\frac{1-\frac{1}{6}\Lambda\,\zeta\bar\zeta}{1+\frac{1}{6}\Lambda\,\zeta\bar\zeta}\right)^2\,\d u\,\d v
 +{\textstyle\frac{1}{3}}\Lambda \left(\frac{1-\frac{1}{6}\Lambda\,\zeta\bar\zeta}{1+\frac{1}{6}\Lambda\,\zeta\bar\zeta}\right)^2v^2\,\d u^2 \ .
\label{MadS}
\ee
For the $\Lambda=0$ case, this is obviously a flat space. For $\Lambda\neq 0$, the parametrization of the
(anti-) de Sitter universe represented geometrically as a hyperboloid $-Z_0^2+Z_1^2+Z_2^2+Z_3^2+\epsilon Z_4^2=3/\Lambda$
($\epsilon\equiv\Lambda/|\Lambda|$) imbedded in a flat five-dimensional space
$\d s^2=-\d Z_0^2+\d Z_1^2+\d Z_2^2+\d Z_3^2+\epsilon\d Z_4^2$ is
\beqn
 Z_0 & = & \frac{1-\frac{1}{6}\Lambda\,\zeta\bar\zeta}{1+\frac{1}{6}\Lambda\,\zeta\bar\zeta}\left[{\textstyle\frac{1}{2}} \left({\textstyle\frac{1}{3}}\Lambda\right)^2 u^2 v + {\textstyle\frac{1}{3}}\Lambda\, u +{\textstyle\frac{1}{2}} \left({\textstyle\frac{1}{3}}\Lambda\right)^{-1} v \right]\ , \nonumber\\
 Z_1 & = & \frac{1-\frac{1}{6}\Lambda\,\zeta\bar\zeta}{1+\frac{1}{6}\Lambda\,\zeta\bar\zeta}\left[{\textstyle\frac{1}{2}} \left({\textstyle\frac{1}{3}}\Lambda\right)^2 u^2 v + {\textstyle\frac{1}{3}}\Lambda\, u -{\textstyle\frac{1}{2}} \left({\textstyle\frac{1}{3}}\Lambda\right)^{-1} v \right]\ , \nonumber\\
 Z_2 & = & {\textstyle\frac{1}{\sqrt{2}}}\frac{\zeta+\bar\zeta}{1+\frac{1}{6}\Lambda\,\zeta\bar\zeta}\ , \qquad Z_3={\textstyle\frac{\hbox{i}}{\sqrt{2}}}\frac{\zeta-\bar\zeta}{1+\frac{1}{6}\Lambda\,\zeta\bar\zeta}\ , \\
 Z_4 & = & \left({\textstyle\frac{1}{3}}|\Lambda|\right)^{-1/2}\,\frac{1-\frac{1}{6}\Lambda\,\zeta\bar\zeta}{1+\frac{1}{6}\Lambda\,\zeta\bar\zeta}\left(1 + {\textstyle\frac{1}{3}}\Lambda\, uv\right) \nonumber \ .
\eeqn
Analogous parameterizations can be found for other (canonical)
choices of $\varepsilon$ and $C$.

\subsection{Type $D$ spacetimes}

In this case, the vanishing of $\Psi_4$ requires $H=H_0$, and the curvature scalars (\ref{psi2})--(\ref{phi22})
reduce to
\beqn
 \Psi_2 & = & -{\textstyle\frac{1}{6}}\left(D-D_0\right)\frac{P^2}{Q^2}\ , \label{Psi2D}\\
  R & = & 24\alpha-12\e(\alpha+\beta)\frac{P}{Q}+2\left(D-D_0\right)\frac{P^2}{Q^2}\ , \label{RD}\\
 \Phi_{11}  & = & {\textstyle\frac{1}{2}\e}(\alpha+\beta)\,\frac{P}{Q}-{\textstyle\frac{1}{4}}\left(D-D_0\right)\,\frac{P^2}{Q^2} \label{Pshi11D}\ , \\
 \Phi_{22}  & = & {\textstyle\frac{1}{2}}\bigg\{(\alpha A_0+A_2)\,Q-\e(\alpha+\beta)(A_0+A_1\zeta+\bar{A}_1\bar\zeta+A_2\zeta\bar\zeta) \nonumber \\
    & & \hskip 20mm{}-2\e(\alpha+\beta)\,v\,\,(\dot{{C}}\,\zeta+\dot{\bar{C}}\,\bar\zeta)\bigg\}\,\frac{P^3}{Q^4}\ . \label{Phi22D}
\eeqn
Notice that the $\Phi_{22}$ component (\ref{Phi22D}) has the same form as in the conformally flat case (\ref{Phi22con})
but the other scalars differ since $D\not=D_0$.

A {\em pure radiation} field requires $\Phi_{11}=0$, which implies that $\>\alpha=\beta={1\over2}\Lambda\>$, $\e=1$,
$C=0$, $D=\Lambda$, and the metric takes the form
\be
 \d s^2={2\,\d\zeta\,\d\bar\zeta\over \left(1+{\textstyle\frac{1}{2}}\Lambda\zeta\bar\zeta\right)^2}
 -2\,\d u\,\d v+\left(\Lambda v^2-H_0\right)\,\d u^2\ ,
 \label{nariaipure}
\ee
the scalars being  $\>\Psi_2 = -{1\over3}\Lambda\>$, $\>R=4\Lambda\>$, and
\be
 \Phi_{22}   =  {\textstyle\frac{1}{2}}\left(A_2-{\textstyle {1\over2}}\Lambda\, A_0\right)
 \left(1-{\textstyle {1\over2}}\Lambda\,\zeta\bar\zeta\right)
 -{\textstyle {1\over2}}\Lambda\,(A_1\zeta+\bar A_1\bar\zeta)\ . \label{phi22pureD}
\ee
When $\Lambda>0$ ($\Lambda<0$) this solution represents the (anti-)Nariai universe \cite{Nariai51}
with pure radiation which can not be an electromagnetic field
(see the general theorem in \cite{VandenBergh89}). However, for $\Lambda=0$ this
becomes the conformally flat electromagnetic plane wave mentioned in
the previous section. Note also that the metric
(\ref{nariaipure}) is identical to the Kundt canonical form (\ref{metricK}) since $P=Q$ in the case
of pure radiation solutions, and consequently $v=w$, $W=0$, ${\cal H}=-{1\over2}\Lambda w^2+{1\over2}H_0$.
In particular, it belongs to the invariant subclass defined by $W_{,w}=0$, and thus seems to be a
counterexample to a conjecture of \cite{WilVan90} that the solutions exhibited there
(for which $W_{,w}\neq 0$) are the only type $D$ pure radiation metrics of the Kundt class.

The situation in which there is no pure radiation ($\Phi_{22}=0$), but {\em only
the $\Phi_{11}$ component} of the matter field, admits two
possibilities. If $\>\e(\alpha+\beta)\neq0\>$ then
$\e=1$, $C=\gamma$, $A_2=\beta A_0$, and $A_1=\gamma A_0$ so that $H=A_0 Q/P$. As in the conformally
 flat case, this function $H$ can always be
removed by a coordinate transformation. Thus, all the
solutions of this subclass can be written as the metric (\ref{metric}) in which
\be
 P  =  1+\alpha\,\zeta\bar\zeta\ ,\qquad
 Q  =  1+\beta\,\zeta\bar\zeta+\gamma\,\zeta+\bar{\gamma}\,\bar\zeta\  ,\qquad
 F  =  D(u)\,{Q^2\over P^2}\,v^2\ ,  \label{coeff2D}
\ee
With $v=r/[\,1+U(u)\,r\,]$, where $U(u)=-{1\over2}\int D(u)\,du$,
we put the solution into the form
\be
 \d s^2={2\,\d\zeta\,\d\bar\zeta\over P^2}-{Q^2\over P^2}\,{2\,\d u\,\d r\over
 [\,1+U(u)\,r\,]^2}\ ,
 \label{metricBRD}
\ee
(For $D=D_0$ the metric (\ref{coeff2D}) reduces to the
conformally flat spacetime (\ref{coeff2}), i.e. (\ref{metricBRD}) gives (\ref{metricBR}).)
Looking for the electrovacuum solutions we require  $R=4\Lambda$ which implies
$\alpha=\beta$, $\gamma=0$, $D=2(\Lambda-\alpha)$, and the scalars are
$\Psi_2=-{1\over3}\Lambda$, $\Phi_{11}=\alpha-{1\over2}\Lambda$.
In such a case, the metric (\ref{metric}) simplifies considerably to
\be
 \d s^2={2\,\d\zeta\,\d\bar\zeta\over (1+\alpha\,\zeta\bar\zeta)^2}-2\,\d u\,\d v
 +2(\Lambda-\alpha)\,v^2\,\d  u^2\ .
 \label{product}
\ee
These are well-known electrovacuum spacetimes with the geometry of a direct product of two constant curvature
2-spaces \cite{Bertotti59}, see also  \cite{CahDef68, kramerbook}. This is obvious from the  form (\ref{metricBRD}) since $P=Q\,$, $\>U(u)=-{1\over2}Du\>$.
Of course, for $\Lambda=0$ we recover the conformally flat solution (\ref{br}).
The second possibility for the case $\Phi_{22}=0$ is $\>\e(\alpha+\beta)=0\>$ which implies  $\>A_2=-\alpha A_0\>$, i.e. $\>H_0=[(1-\alpha\,\zeta\bar\zeta)A_0+A_1\zeta+\bar{A}_1\bar\zeta]/P$.
The corresponding scalars are
$ R  =  24\alpha+2(D-D_0)P^2/Q^2$, where $D_0=2\e\alpha+2C\bar C$,
$ \Psi_2  =  2\alpha-{\textstyle\frac{1}{12}}R$,
$ \Phi_{11}  =  3\alpha-{\textstyle\frac{1}{8}}R$.
For electrovacuum solutions, $R=4\Lambda$ so that $\alpha=0=\beta$, $\varepsilon=1$, $C=0$, $D=2\Lambda<0$.
We thus uniquely obtain
the exceptional Pleba\'nski--Hacyan universe \cite{PlebHac79}
\be
 \d s^2=2\,\d\zeta\,\d\bar\zeta-2\,\d u\,\d v
 +\left(2\Lambda\,v^2 -A_1\zeta-\bar{A}_1\bar\zeta \right)\,\d  u^2\ ,
 \label{PB}
\ee
for which  $\Phi_1=\sqrt{-\frac{1}{2}\Lambda}\,e^{i \varphi}$ (the function $A_0$ can be removed by a coordinate
transformation \cite{PlebHac79}). When $A_1=0$, this reduces to the form (\ref{product}) of the direct product spacetimes.

Again, type $D$ spacetimes  which combine $\Phi_{11}$ and  $\Phi_{22}$ can be considered. For example,
there exist solutions which  represent pure radiation in the  electrovacuum universes (\ref{product}) and
(\ref{PB}), see e.g. \cite{OrtPod02}.

Finally, the type $D$ Einstein spaces with {\em no matter} are given by (\ref{nariaipure})-(\ref{phi22pureD})
with the constraints $A_2=\frac{1}{2}\Lambda A_0$, $A_1=0$. In such a case, the function $H_0=A_0$
can again be removed by a suitable coordinate transformation. Thus, without
loss of generality one obtains
\be
 \d s^2={2\,\d\zeta\,\d\bar\zeta\over \left(1+{\textstyle\frac{1}{2}}\Lambda\zeta\bar\zeta\right)^2}
 -2\,\d u\,\d v+\Lambda v^2\,\d u^2\ ,
 \label{nariai}
\ee
which is the (anti-)Nariai universe \cite{Nariai51}, discussed recently in \cite{Ortaggio02}. Obviously, this is also the direct product spacetime  (\ref{product}) for $\alpha={1\over2}\Lambda$.
There is no type $D$ vacuum solution of the form (\ref{metric}) for $\Lambda=0$ since (\ref{nariai}) reduces to Minkowski space in such a case.

The main results presented above are summarized in the table 2.

\begin{table}[htbp]
 \[
  \begin{array}{|c||c|c|} \hline
    & &   \\
    & \mbox{type } O & \mbox{type } D \\
    & &  \\ \hline\hline
    & \mbox{ Minkowski } (\Lambda=0) & \longleftarrow\quad \mbox{ none for } \Lambda=0  \\
   \mbox{ no matter} & \mbox{ de Sitter } (\Lambda>0) & \mbox{ Nariai } (\Lambda>0) \\
   \Phi_{11}=0, \Phi_{22}=0 &  \mbox{ anti-de Sitter } (\Lambda<0)  & \mbox{ anti-Nariai } (\Lambda<0) \\ \hline
    & \mbox{ Baldwin--Jeffery--Brdi\v cka } (\Lambda=0) & \longleftarrow\quad \mbox{ none for } \Lambda=0 \\
   \mbox{ pure radiation } & \mbox{ Edgar--Ludwig } (\Lambda=0) &  \\
   \Phi_{11}=0, \Phi_{22}\neq 0 & \mbox{ Ozsv\'ath--Robinson--R\'ozga } (\Lambda\neq 0)
     & \ (\ref{nariaipure}) \mbox{ for } \Lambda\ne0 \\ \hline
    &  (\ref{coeff2}), (\ref{metricBR}) & \ (\ref{coeff2D}), (\ref{metricBRD})   \\
   \mbox{ no pure radiation } &  \mbox{ Bertotti--Robinson } (\Lambda=0)  &  \mbox{ direct product spacetimes} \\
   \Phi_{11}\neq 0, \Phi_{22}=0 &  & \mbox{ Pleba\'nski--Hacyan } (\Lambda<0) \\ \hline
    &  &  \\
   \mbox{ general } &  \ (\ref{metric}) \hbox{ with $H=H_0$, $D=D_0$ } & \ (\ref{metric})  \hbox{ with $H=H_0$, $D\neq D_0$ }   \\
   \Phi_{11}\neq 0, \Phi_{22}\neq 0 &    &   \\ \hline
  \end{array}
 \]
 \caption{Summary of the background spacetimes and the most important subcases.}
 \label{backgrounds}
\end{table}


\section{Exact gravitational waves of arbitrary profile on the above backgrounds}

\label{sec_waves}

By considering an arbitrary function $H(\zeta,\bar\zeta,u)$ in the metric (\ref{metric}), (\ref{coeff}), different from
$H_0$ as introduced in (\ref{H0}), the  scalar  $\Psi_4$  representing gravitational radiation becomes non-vanishing,
see (\ref{psi4}). As the coordinate $u$ plays the role of the retarded time, {\it gravitational waves of arbitrary
profiles} $\ H^{w}\equiv H-H_0\ $
can thus be introduced into the above spacetimes. When the backgrounds are taken to be conformally flat,
exact radiative spacetimes of Petrov type $N$ are obtained. For the backgrounds of type $D$ one gets explicit
gravitational waves of type $II$, see table~1. It is obvious from (\ref{metric}) that the wave-fronts $\>u=\>$const.
are \emph{non-expanding} 2-spaces with a \emph{constant curvature} equal to $2\alpha$.

It can be observed  from (\ref{psi2})--(\ref{phi22}) that the function $H$ does not appear in the scalars $\Psi_2$, $R$, and $\Phi_{11}$. In fact, introducing the additional term $\>{-{(Q/P)\,H^{w}\,\d u^2}}\>$ in the metric corresponds to a generalised Kerr--Schild transformation (see e.g. \cite{Taub81,Bal00}) of the background geometries\footnote{The authors are grateful to the referee for this observation.}.
Consequently, the radiative spacetimes (\ref{metric}) of  type $N$ and $II$ have {\it exactly the
same values} of the Weyl component $\Psi_2$, the Ricci curvature $R$, and the non-null matter component $\Phi_{11}$
as the corresponding  backgrounds described in the previous section (see table~2). The only difference, apart from
the introduction of the $\Psi_4$ component, may be in
the pure radiation field. Indeed, the component $\Phi_{22}$, which is {\em linear} in $H$, can now be understood as
a superposition of the background term $\Phi_{22}^{b}$ given by  (\ref{Phi22con}), identical to (\ref{Phi22D}),
with the term $\Phi_{22}^{w}$ related to the presence of  gravitational radiation,
\be
 \Phi_{22}^{w}= {\textstyle\frac{1}{2}}\bigg\{\Big[\,\e(\alpha-\beta)\,(1-\alpha\,\zeta\bar\zeta)\,
+2\alpha\,(C\,\zeta+\bar{C}\,\bar\zeta)\,\Big]\,H^{w}+P^2Q\,H^{w}_{,\zeta\bar\zeta}
\bigg\}\,\frac{P^3}{Q^4}\ .\label{Hw}
\ee
In general, a gravitational wave ($\Psi_4\ne0$) is thus accompanied by the (additional) pure radiation  field $\Phi_{22}^{w}$.

However, in special situations when $\Phi_{22}^{w}=0$, the gravitational wave is {\it not} related to the above pure radiation. For example when $P=Q=1+\alpha\,\zeta\bar\zeta$, i.e.
for spacetimes (\ref{product}) which are direct product of two
constant curvature 2-spaces (Minkowski, Bertotti--Robinson, (anti-)Nariai, Pleba\'nski--Hacyan spaces),
the relation $\Phi_{22}^{w}=0$ is satisfied if and only if $H^{w}=f(\zeta,u)+\bar f(\bar\zeta,u)$.
This is just the Einstein equation for purely gravitational waves propagating in Einstein spaces, in
which case $\Phi_{22}^{b}=0=\Phi_{11}$.

Let us now discuss in some detail the particular subclasses which include gravitational waves.

\subsection{Type $N$ spacetimes}

A complete class of  non-expanding type $N$ {\it vacuum} solutions with (possibly) non-vanishing cosmological constant
was found by Ozsv\'ath, Robinson and R\'ozga \cite{OzsRobRoz85}, and later studied also from a physical point of view \cite{BicPod99II}. These Einstein spaces represent exact  pure gravitational waves which propagate  in  constant curvature backgrounds, i.e. in Minkowski, de~Sitter or anti-de~Sitter universe (for $\Lambda=0$, $\Lambda>0$ or $\Lambda<0$, respectively). For example, considering the background $\d s_{b}^2$ given by  (\ref{MadS}), one obtains radiative solutions of the form
\be
 \d s^2=\d s_{b}^2-
\left(\frac{1-\frac{1}{6}\Lambda\,\zeta\bar\zeta}{1+\frac{1}{6}\Lambda\,\zeta\bar\zeta}\right)
\,H^{w}\,\d u^2\ .
 \label{backpluswave}
\ee
For vanishing cosmological constant we obtain exactly the well-known class of {\it pp\,}-waves
\cite{kramerbook}. Alternatively, with other forms of  constant curvature backgrounds corresponding to different canonical choices of the parameters $\e$ and $C$ other classes are obtained, namely
the specific Kundt spacetimes for $\Lambda=0$ \cite{kundt1, kundt62, EK} or the  Siklos family
\cite{Siklos85} for $\Lambda<0$, see \cite{OzsRobRoz85, BicPod99I}.
When $\Phi_{22}^{w}=0$ these are pure gravitational waves, otherwise they are accompanied by a pure radiation component.

Another possibility is to consider the conformally flat {\it pure radiation} ($\Phi_{22}^b\not=0$, $\Phi_{11}=0$)
backgrounds $\d s_{b}^2$ of the form (\ref{coeff1}) and introduce $H^w$. Again the {\it pp\,}-waves and
gravitational waves in the Edgar--Ludwig type
backgrounds (\ref{ELgen}) with any $\Lambda$ are thus obtained. In fact, these radiative spacetimes are contained
in the  Ozsv\'ath--Robinson--R\'ozga   family \cite{OzsRobRoz85}.

On the other hand, considering the conformally flat backgrounds (\ref{metricBR}) {\it without pure radiation}
($\Phi_{22}^b=0$, $\Phi_{11}\not=0$) gravitational waves (plus possibly pure radiation if $\Phi_{22}^w\not=0$) are generated. In particular, one obtains the type $N$ solutions representing  gravitational radiation in the Bertotti--Robinson electrovacuum universe
\be
 \d s^2=\d s_{b}^2-H^{w}\,\d u^2\ .
 \label{backpluswaveBR}
\ee
where $\d s_{b}^2$ is given by (\ref{br}).

Of course, gravitational waves in more general conformally flat backgrounds (\ref{Rcon})--(\ref{Phi22con}) with
${\Phi_{11}\not=0}$, $\Phi_{22}^b\not=0$ are also possible.

\subsection{Type $II$ spacetimes}

All type $II$ solutions of the form  (\ref{metric}), (\ref{coeff}) which represent exact gravitational waves propagating in backgrounds with the matter field component $\Phi_{11}$ {\it vanishing} can be written in the form
(\ref{backpluswaveBR})  where $\d s_{b}^2$ is now the metric (\ref{nariaipure}). These describe gravitational radiation in the (anti-)Nariai (type $D$) universe  filled with a pure radiation field $\Phi_{22}=\Phi_{22}^{b}+\Phi_{22}^{w}$, where $\Phi_{22}^{b}$ is given by (\ref{phi22pureD}) and
$\Phi_{22}^{w}$  by (\ref{Hw}), which now reduces to a simple expression $\Phi_{22}^{w}={1\over2}(1+{1\over2}\Lambda\zeta\bar\zeta)^3H_{\zeta\bar\zeta}^{w}\,$.

As an important special subcase we easily obtain {\it pure gravitational waves} of the above form by considering,
not only $\Phi_{11}=0$, but also $\Phi_{22}=0$, i.e. Einstein spaces of the Petrov type $II$ without matter.
Obviously, these solutions can explicitly be written as
\be
 \d s^2=\d s_{b}^2-\left[\,f(\zeta,u)+\bar f(\bar\zeta,u)\,\right]\,\d u^2\ ,
 \label{backpluspurewave}
\ee
where $\d s_{b}^2$ is the metric (\ref{nariai}) of the (anti-)Nariai vacuum universe with
$\Lambda>0$ ($\Lambda<0$), and $f(\zeta,u)$ is an arbitrary function (holomorphic in $\zeta$) which characterizes the profile of the gravitational wave.
Note that the solution (\ref{backpluspurewave}) of the Einstein vacuum equations with cosmological constant
 is included in the class of solutions that was investigated from a different point of view by Lewandowski \cite{Lewand92}.

The complementary situation in which $\Phi_{22}^b=0$, but $\Phi_{11}\not=0$, corresponds to introducing
type~$II$ waves into the background (\ref{metricBRD}). In particular, considering only the
{\it non-null electrovacuum} background universes we obtain gravitational waves in the spacetimes $\d s_{b}^2$
which are a direct product of two constant curvature 2-spaces  (\ref{product}).
The second possibility for the electrovacuum case $\Phi_{22}^b=0$ is  the exceptional  Pleba\'nski--Hacyan
background universe  (\ref{PB}). In both these cases the radiative metric has the form (\ref{backpluspurewave})
with the corresponding form of $\d s_{b}^2$.

Again, these gravitational waves are in general accompanied by a pure radiation contribution $\Phi_{22}^w$.
As particular cases of the type~$II$ spacetimes  {\it without pure radiation}, $\Phi_{22}^b=0=\Phi_{22}^w$,
we obtain the special electrovacuum solutions with gravitational waves that were found by Garc\'\i a and Alvarez
\cite{GarAlvar84}. For $\alpha=\beta=C=0$, $\e=1$, $D=2\Lambda$,
this is their special II-$E_{(+)}$ solution with $\Lambda<0$ which can be written in the form (\ref{backpluspurewave})
where $\d s_{b}^2$ is the Pleba\'nski--Hacyan metric (\ref{product}). For $\alpha=\beta=\Lambda$, $C=0$, $\e=1$, $D=0$
we obtain the (non-twisting subclass of) II-$E_{(-)}$ solution with $\Lambda>0$ in the form (\ref{backpluspurewave})
where $\d s_{b}^2$ is the other direct product Pleba\'nski--Hacyan metric of the form (\ref{product}). In both cases, $\Psi_2=-{1\over3}\Lambda$, $R=4\Lambda$, $\Phi_{11}={1\over2}|\Lambda|$.
Note that all the above mentioned metrics representing pure gravitational waves in electromagnetic universes
are in fact specific subcases of a solution presented by Khlebnikov \cite{Khlebnikov86}.

Finally, we note that gravitational waves (plus possibly pure radiation $\Phi_{22}^w$) in general type~$D$
 backgrounds (\ref{Psi2D})--(\ref{Phi22D}) for which both  ${\Phi_{11}\not=0}$ and $\Phi_{22}^b\not=0$ are easy to construct.

\subsection{Electrovacuum  solutions}

\label{subsec_elvac}

It may  be useful to summarize all the possible cases in which the solution of the form
(\ref{metric}), (\ref{coeff}) represents a spacetime containing an {\it electromagnetic field but
no other matter fields}.
In section~\ref{sec_backgrounds} we concentrated on the background spaces
of type $O$ and $D$. We now investigate the corresponding situations
when gravitational waves are present, i.e. the spacetimes of type $N$ or $II$, respectively.

We have demonstrated (see also the appendix) that the only {\it conformally flat}
spacetimes (including $\Lambda$)
which satisfy the Einstein--Maxwell equations are some special {\it pp\,}-waves with a null
Maxwell field, and the Bertotti--Robinson universe with a non-null Maxwell field. For {\it type}~$N$
{\it pp\,}-waves, corresponding to $P=1=Q$, $F=-H$ in (\ref{metric}), it is well known \cite{kramerbook}
that there is a general combination of electromagnetic and gravitational waves when
$H=2\,|\int E_2\,\d\zeta|^2+f(u,\zeta)+\bar f(u,\bar\zeta)$, where $E_2(u,\zeta)$ represents the
null Maxwell field (\ref{E2}). On the other hand, one has type~$N$ gravitational waves propagating in the
Bertotti--Robinson electrovacuum universe, with $P=1+\alpha\zeta\bar\zeta=Q$, $F=-2\alpha v^2-H$
in (\ref{metric}), for $H=f(u,\zeta)+\bar f(u,\bar\zeta)$, cf. \cite{Khlebnikov86,Ortaggio02,OrtPod02}.

For {\it type} $D$ solutions we may have only a non-null Maxwell field, represented by spacetimes which
are the direct product of two constant curvature 2-spaces (\ref{product}) plus the exceptional
Pleba\'nski--Hacyan universe (\ref{PB}). Pure {\it type} $II$ gravitational waves  in such
type~$D$ electrovacuum spacetimes arise for  $H=f(u,\zeta)+\bar f(u,\bar\zeta)$,
see \cite{GarAlvar84,Khlebnikov86,Ortaggio02,OrtPod02}.

Finally, we consider {\it vacuum backgrounds} and possible {\it electromagnetic waves propagating on these}.
Conformally flat vacuum spacetimes are just spaces of constant curvature. Electromagnetic waves in
the Minkowski spacetime are simply  the {\it pp\,}-waves discussed above. Electromagnetic waves in
the (anti-)de Sitter space were discussed in \cite{OzsRobRoz85}, and are necessarily accompanied
by gravitational waves (see \cite{OzsRobRoz85} for explicit formulae). The only remaining
possibility is thus given by waves in vacuum backgrounds of type $D$, i.e. in the (anti-)Nariai universe
(\ref{nariai}). Interestingly, also in this case the Einstein--Maxwell equation (\ref{phis}),
(\ref{Hw}) for the profile
function $H$ can be integrated explicitly. Exactly as in the case of \pp waves, this admits the
general solution $H=2\,|\int E_2\,\d\zeta|^2+f(u,\zeta)+\bar f(u,\bar\zeta)$. However,
these electromagnetic waves are now necessarily accompanied by a gravitational wave component
(see section~\ref{sec_backgrounds} and \cite{VandenBergh89}). Such an explicit solution for
{\em electromagnetic (plus gravitational) waves in the (anti-)Nariai universe} seems to have
remained unnoticed in the literature so far (whereas pure gravitational waves were already
considered \cite{Khlebnikov86,Ortaggio02}).

\subsection{Effects on test particles}

In order to analyze the effects of the gravitational and matter fields of the above solutions, it is natural
to investigate the specific influence of  various components of these fields on the relative motion of free test particles.
Such a local characterization of spacetimes, based on the equation of geodesic deviation, was described
in the pioneering works by Pirani \cite{piraniriem, piraniinvariant} and Szekeres \cite{szekerescompass, szekeresmatter}.
Following the notation introduced in \cite{BicPod99II} we set up an orthonormal frame
 $\{ \E_{(a)} \}$, $\E_{(a)} \cdot \E_{(b)} =\eta_{(a)(b)}\equiv\hbox{diag}(-1,1,1,1)$,
such that  the timelike vector $\E_{(0)}$ coincides at a given event with the four-velocity $\uu$
of a geodesic test observer.
By projecting the equation of geodesic deviation onto this frame we obtain
\be
\ddot Z^{(i)}=-R^{(i)}_{\ (0)(j)(0)}Z^{(j)}\ , \label{geodev}
\ee
where \ $i, j=1,2,3$. The  frame components of the displacement
vector $Z^{(j)}\equiv e^{(j)}_\mu Z^\mu$ determine directly the
distance between close test particles, and
$\ddot Z^{(i)}\equiv e^{(i)}_\mu({D^2Z^\mu}/{d\tau^2})$
are their physical relative accelerations. From the standard decomposition of the curvature tensor
 (see, e.g., Eqs. (3.44)--(3.47) in \cite{kramerbook}), we immediately obtain
$R_{(i)(0)(j)(0)}=C_{(i)(0)(j)(0)}+{1\over2}(\,\delta_{ij}S_{(0)(0)}-
S_{(i)(j)}\,)+{1\over12}R\,\delta_{ij}$, where $C_{(i)(0)(j)(0)}$ are components of the Weyl tensor
and  $R$ and $S_{(a)(b)}$ respectively denote the trace and the traceless part of the Ricci tensor.
These frame components can conveniently be expressed using the corresponding null complex tetrad
\be
\m    ={\textstyle{1\over\sqrt 2}}\left(\E_{(1)}+\hbox{i}\, \E_{(2)}\right)\ ,\qquad
\bl ={\textstyle{1\over\sqrt 2}}\left(\E_{(0)}-\E_{(3)}\right)    \ ,\quad
\k  ={\textstyle{1\over\sqrt 2}}\left(\E_{(0)}+\E_{(3)}\right)\ .\label{E2.9}
\ee
In the natural tetrad (\ref{privtetrad}), the only non-vanishing scalars are those given by
(\ref{psi2})--(\ref{phi22}) so that
\beqn
&&C_{(1)(0)(1)(0)}= {\textstyle{1\over2}}\,\R e\,\Psi_4 -\Psi_2  \ ,\quad
C_{(2)(0)(2)(0)}= -{\textstyle{1\over2}}\,\R e\,\Psi_4 -\Psi_2  \ ,\nonumber\\
&&C_{(1)(0)(2)(0)}=-{\textstyle{1\over2}}\,\I m\,\Psi_4
\ ,\quad C_{(3)(0)(3)(0)}=2\,\Psi_2  \ ,\label{CC}\\
&&S_{(0)(0)}= \Phi_{22}+ 2\,\Phi_{11} \ ,\quad
S_{(1)(1)}= 2\,\Phi_{11}  \ ,\nonumber\\
&&S_{(3)(3)}=\Phi_{22}- 2\,\Phi_{11} \ ,\quad
S_{(2)(2)}=2\,\Phi_{11}  \ .\nonumber
\eeqn
Therefore, the equation of geodesic deviation (\ref{geodev}) takes the form
\beqn
\ddot Z^{(1)}&=&{\textstyle{1\over12}}R\,Z^{(1)}\ -\
               {\textstyle{1\over2}}\Phi_{22}\,Z^{(1)}\ +\
               \C Z^{(1)}\ -\ \A_+\,Z^{(1)}+\A_\times\,Z^{(2)} \ ,\nonumber\\
\ddot Z^{(2)}&=&{\textstyle{1\over12}}R\,Z^{(2)}\ -\
               {\textstyle{1\over2}}\Phi_{22}\,Z^{(2)}\ +\
               \C Z^{(2)}\ +\ \A_+\,Z^{(2)}+\A_\times\,Z^{(1)}
                        \ ,\label{geodevi}\\
\ddot Z^{(3)}&=&{\textstyle{1\over12}}R\,Z^{(3)}\ -\
               2\Phi_{11}\,Z^{(3)}\ - 2\C Z^{(3)}
               \ ,\nonumber
\eeqn
where
\be
\C=\Psi_2\ ,\qquad
\A_+={\textstyle{1\over2}}\,\R e\,\Psi_4\ ,\qquad
\A_\times={\textstyle{1\over2}}\,\I m\,\Psi_4\ .\label{coeffi}
\ee

Equations (\ref{geodevi}) are well suited for physical
interpretation. Clearly, the relative motions of nearby test particles depend on:

\begin{enumerate}
\item  the Ricci scalar $R$, which is responsible for overall background
   {\it isotropic}  motions (in the case of pure radiation or vacuum spacetimes
   this is a constant factor proportional to the cosmological constant,
   ${1\over12}R={1\over3}\Lambda$) ;
\item  the terms describing the matter-content which consist of two different components:
   the {\it pure radiation} field $\Phi_{22}$ affects the motion only in the {\it transverse}
   plane spanned by the vectors $\E_{(1)}, \E_{(2)}$, whereas the {\it non-null} field influences
   only the {\it longitudinal} direction $\E_{(3)}$;
\item  the terms which depend on the local free gravitational field. There is a {\it Coulomb} component with the
   amplitude $\C$ for type $D$ and $II$ spacetimes, and transverse
   components with amplitudes given by $\A_+$ and $\A_\times$ representing the effect of
   {\it gravitational waves} on the particles in type $N$ and $II$ spacetimes.
\end{enumerate}

It can be observed that gravitational radiation is present if and only if $H\not=H_0$.
The corresponding  effects are typical for both linearized   and exact gravitational waves
(cf. e.g. \cite{MTW, szekerescompass, BicPod99II}). There are two polarization modes ``+''  and  ``$\times$''
of a transverse gravitational wave with amplitudes $\A_+$ and $\A_\times$.
A ring of test particles is deformed into an ellipse in the plane perpendicular to the direction
of propagation, the axes of different polarizations being shifted one with respect
to the other by ${\pi\over4}$. The structure of the equations (\ref{geodevi}) thus supports our interpretation of
the solutions (\ref{metric}) as exact gravitational waves which propagate in various background universes.
Notice that the coefficients $R$, $\Phi_{11}$ and $\Psi_2$ remain the
same as for the corresponding background, whereas the amplitudes $\Phi_{22}$ and
$\Psi_4$ which represent radiation depend on the specific form of the
structural function $H$.

The above analysis only applies to the reference frame
(\ref{E2.9}) related to the natural null tetrad
(\ref{privtetrad}). However, this can easily be generalized to a
case in which the orthonormal interpretation frame is
adapted to an {\it arbitrary observer} passing along a timelike geodesic through
the given event. Performing the Lorentz transformation
\beqn
&&\k'  =A^{-1}\,\k     \ ,\nonumber\\
&&\bl' =A\,\bl+B\,\bar\m+\bar B\,\m+B\bar B A^{-1}\,\k \ , \label{Lorentz}\\
&&\m'  =\m+BA^{-1}\,\k\ ,\nonumber
\eeqn
where $A=\sqrt2\,(Q^2/P^2)\,\dot u\,$, $\,B=\sqrt2\,(1/P)\,\dot\zeta$, we obtain
from (\ref{privtetrad}) the null tetrad
\beqn
 {\bf m'} & = & \frac{P\,\dot\zeta}{Q^2\,\dot u}\,\pa_v+P\,\pa_{\bar{\zeta}}\ ,  \qquad
 {\bf\bar m'}  =  \frac{P\,\dot{\bar\zeta}}{Q^2\,\dot u}\,\pa_v+P\,\pa_\zeta\ , \nonumber \\
 {\bf l'} & = & \Big(\sqrt2\,\dot v-\frac{P^2}{\sqrt2\, Q^2\,\dot u}\,\Big)\,\pa_v
   +\sqrt2\,\dot \zeta\,\pa_\zeta+\sqrt2\,\dot{\bar\zeta}\,\pa_{\bar\zeta}
   +\sqrt2\,\dot u\,\pa_u\ , \label{gentetrad}\\
 {\bf k'} & = & \frac{P^2}{\sqrt2 \, Q^2\,\dot u }\,\pa_v\ . \nonumber
\eeqn
Obviously, $\E_{(0)}' \equiv{\textstyle{1\over\sqrt 2}}\left(\bl'+\k'\right)=
\dot v\,\pa_v +\dot \zeta\,\pa_\zeta+\dot{\bar\zeta}\,\pa_{\bar\zeta}+\dot
u\,\pa_u$ represents the local value of the four-velocity of a general geodesic observer
(with the dot indicating a differentiation with respect to its proper time)
normalized as
\be
\E_{(0)}'\cdot\E_{(0)}'=2\,{1\over P^2}\,\dot\zeta\,\dot{\bar\zeta}
   -2\,{Q^2\over P^2}\,\dot u\,\dot v+F\,\dot  u^2=-1\ .
 \label{normal}
\ee
After the Lorentz transformation (\ref{Lorentz}), the corresponding
(non-vanishing) coefficients take the form
\beqn
&&\Psi_2'=\Psi_2\ ,\qquad
  \Psi_3'=3\bar B\,\Psi_2\ ,\qquad
  \Psi_4'=A^2\Psi_4+6\bar B^2\,\Psi_2   \ ,\label{Psi'}\\
&&\Phi_{11}'=\Phi_{11} \ ,\qquad
  \Phi_{12}'=2B\,\Phi_{11}\ ,\qquad
  \Phi_{22}'=A^2\,\Phi_{22}+4B\bar B\,\Phi_{11}\ .\nonumber
\eeqn
Therefore, the effect on the amplitudes introduced in (\ref{geodevi}), (\ref{coeffi}) is such that
the Coulomb components $\C$ and $\Phi_{11}$ of the fields
remain the same, whereas the amplitudes $\A$ and $\Phi_{22}$ representing gravitational
and pure radiation measured by the corresponding observer in the frame  $\{ \E_{(a)}' \}$
are rescaled by the factor $A^2$. Moreover,
for geodesics with $\dot\zeta\not=0$ there are  ``kinematic'' contributions to $\A$
and $\Phi_{22}$  which involve  $B$. Additional terms $\Psi_3'$ and $\Phi_{12}'$
proportional to $B$ also enter the equation of geodesic deviation (\ref{geodev}) in
such a case through the contributions
\beqn
&&C_{(1)(0)(3)(0)}'=\R e\,\Psi_3'
\ ,\quad C_{(2)(0)(3)(0)}'=-\I m\,\Psi_3'  \ ,\nonumber\\
&&\quad S_{(1)(3)}'= -2\,\R e\,\Phi_{12}'\ ,\quad
S_{(2)(3)}'= -2\,\I m\,\Phi_{12}'  \ .\label{CC'SS'}
\eeqn
These are responsible for effects which relate the displacement in the longitudinal direction
$\E_{(3)}'$ to acceleration in the transverse directions $\E_{(1)}'$, $\E_{(2)}'$, and vice versa.

Note finally (cf. the general analysis  in \cite{BicPod99II})
that the interpretation frame (\ref{gentetrad}) is \emph{parallelly
transported} along the  geodesic if and only if
$(Q/P)_{\bar\zeta}=0=(Q/P)_\zeta$. This condition is equivalent to
vanishing of the spin coefficients $\tau=0=\pi$, i.e. to the
choice $P=Q=1+\alpha\,\zeta\bar\zeta$. Therefore, the tetrad is
parallelly transported along all timelike geodesics in the
spacetimes of the form (\ref{product}) which are direct product of two
constant curvature 2-spaces, in particular in Minkowski,
Bertotti--Robinson, (anti-)Nariai, Pleba\'nski--Hacyan spaces, and
in all the corresponding spacetimes discussed in this section, representing
radiation on these backgrounds.

\section{Concluding remarks}

An explicit family  (\ref{metric}),  (\ref{coeff}) of the Kundt non-diverging class has
been presented which depends on three constant parameters $\alpha$, $\beta$, $\e$
and three structural functions  $C(u)$, $D(u)$  and $H(\zeta,\bar\zeta,u)$. This contains many
particular solutions of  Petrov types $O$, $D$, $N$ and $II$ which, for specific choices
of the constants and functions, admit a cosmological constant,
electromagnetic field, pure radiation and/or general non-null matter component.
Among these are, for example, the conformally flat  Minkowski and (anti-)de Sitter
Einstein spaces, the Edgar--Ludwig pure radiation solutions or the Bertotti--Robinson electrovacuum
universe. The type $D$ solutions contain the (anti-)Nariai, Pleba\'nski--Hacyan or other
direct product electrovacuum spacetimes, and their generalizations (see section~\ref{sec_backgrounds}
and table~2). Moreover, all these $O$ and $D$ solutions can naturally be understood as ``backgrounds'' for
the corresponding type $N$ and $II$ radiative spacetimes, respectively. As described in
section~\ref{sec_waves}, this is  achieved simply by considering an additional term
$\>-(Q/P)\,H^w\,\d u^2\>$ in the metric, which is
proportional to the function $H^w(\zeta,\bar\zeta,u)$. Since the coordinate $u$  is the
retarded time, exact gravitational waves of {\it arbitrary profiles} which propagate in the corresponding
background universes can thus easily be introduced.

In particular, it is straightforward to
construct {\it sandwich waves} by considering the function of the form $H^w=h(\zeta,\bar\zeta)\,d(u)$,
where the profile  $d(u)$ is non-vanishing only on some finite interval representing
the wave-zone between two non-radiative background parts in front and behind the wave.
Of course, the background need not be flat as in the case of well-known sandwich {\it pp\,}-waves
\cite{BPR}, but can be of {\it any} kind summarized in table 2 --- the universe  in which the sandwich wave
propagates can be curved and filled with various matter contents, such as the electromagnetic field and/or
pure radiation.

Also, it is possible to construct {\it impulsive limits} of such sandwich waves by considering the profile
$d(u)$ to approach the Dirac delta distribution $\delta(u)$. When the background is Minkowski or (anti-)de~Sitter universe, non-expanding impulsive gravitational waves in spacetimes of constant curvature are recovered,
see e.g. \cite{Podolsky02} for a review. Impulsive waves of this type in the Nariai universe
have recently been found and studied  in \cite{Ortaggio02}, and in all direct product spacetimes
(including anti-Nariai and Bertotti--Robinson \cite{Ortaggio02}) in \cite{OrtPod02}.

The family of Kundt's solutions presented in this paper thus represents an interesting
generalization of some previously known but ``separated'' results. Its main advantage is that
various particular background and radiative spacetimes are described using a unified, simple
and explicit metric form.
Of course, most of the special subclasses are well-known. However, we have discussed here all the possible
Petrov types and matter contents systematically, identifying some spacetimes which have been
overlooked in the literature so far. For example, the type~$D$ solution  (\ref{nariaipure})
representing pure radiation in (anti-)Nariai universe, which seems to be a counterexample to the
conjecture proposed in \cite{WilVan90}, has been found. In subsection~\ref{subsec_elvac} we have
also presented a new explicit type $II$ electrovacuum solution which describes electromagnetic and
gravitational waves propagating in the (anti-)Nariai background.

Finally, it is worth mentioning that some of the solutions
presented above belong to the recently obtained class of
spacetimes for which all of the scalar curvature invariants vanish
\cite{PraPraColMil02}. They may play an important role since these are exact
solutions in string theory to all perturbative orders \cite{Coley02}. In fact,
{\it all\,} type~$N$ and type~$O$ spacetimes with vanishing
invariants with $\Phi_{12}=0$ are contained in our family of
solutions  (\ref{metric}),  (\ref{coeff}).

\section*{Acknowledgments}

We wish to thank Professor Hans Stephani for providing us with some recent references related
to the topic. The work was supported by  the grant GACR-202/02/0735 from the Czech Republic,
by Charles University in Prague and by INFN.

\section*{Appendix. Conformally flat Einstein--Maxwell fields with $\Lambda$}
\renewcommand{\theequation}{A\arabic{equation}}
\setcounter{equation}{0}

All conformally flat electrovac metrics with a vanishing cosmological constant $\Lambda$
have been known for a long time \cite{CahLer66,Stephani67,TarTup74,McLTarTup75}.
Demanding a spacetime to have a vanishing Weyl tensor and to satisfy the Einstein--Maxwell
equations is so a severe restriction that one is left with only two possibilities,
according to the algebraic type of the Maxwell field. For a \emph{non-null} Maxwell
field one obtains the Bertotti--Robinson \cite{LeviCivita17BR,Bertotti59,Robinson59}
universe. This is given by the metric (\ref{br}) or, alternatively, in a different coordinate system such that
$\zeta=\sqrt{2}\,a\,e^{i\phi}\tan(\theta/2)$, $a^2=1/2\alpha$,
\beqn
t&=&\frac{a(-vu^{2}+2ua^{2}+2va^{2})}{-vu^{2}+2ua^{2}-2va^{2}-2\sqrt{2}\,avu +
2\sqrt{2}\,a^{3}}\ , \nonumber\\
r&=&\frac{2\sqrt{2}\,a^{4}}{-vu^{2}+2ua^{2}-2va^{2}-2\sqrt{2}avu+2\sqrt{2}\,a^{3}}\ , \nonumber
\eeqn
it takes form
\be
 \d s^2=\frac{a^2}{r^2}\,[\,-\d t^2+\d r^2+r^2(\d\theta^2+\sin^2\theta\,\d\phi^2)\,] \ ,
 \label{BR}
\ee
which is explicitly conformally flat. Alternatively, for a \emph{null} field the solution is
represented by the special \pp wave \cite{BalJef26,Brdicka51}
discussed in section \ref{sub_confflat}
\be
 \d s^2=2\,\d\zeta\d\bar\zeta-2\,\d u\d v-A_2(u)\zeta\bar\zeta\,\d u^2 \ ,
 \label{plane}
\ee
where $A_2(u)$ is an arbitrary (positive) function.

Here we consider a more general problem by admitting $\Lambda\neq0$, i.e. to find all possible conformally
flat solutions of the Einstein--Maxwell system
\beqn
 & & R_{ab}-\textstyle{\frac{1}{2}}Rg_{ab}+\Lambda g_{ab}=T_{ab}
     =\textstyle{\frac{1}{2}}F^{\ c}_a\bar{F}_{bc} \ ,   \nonumber \\
 & & F^{ab}_{\quad ;\,b}=0 \ , \label{EM}
\eeqn
where $F_{ab}$ denotes the complex self-dual Maxwell tensor. We prove, in fact, that any such solution
necessarily requires $\Lambda=0$, and thus it reduces to either (\ref{BR}) or (\ref{plane}).
Note that this was already asserted by Khlebnikov and Shelkovenko
\cite{KhlShe76} but their result seems to have been somewhat overlooked.
Moreover, they did not present an explicit proof.
We employ the Newman--Penrose (NP) formalism, and refer to \cite{kramerbook} for the complete set of
general equations. Of course, conformal flatness is equivalent to setting all components of the Weyl tensor
to zero, $\Psi_i=0$ for $i=0,1,2,3,4$, whereas the tracelessness of $T_{ab}$ requires $R=4\Lambda=\mbox{constant}$.
We now separately treat the two possible cases of non-null and null Maxwell fields.

\subsection*{Non-null fields}

Given an arbitrary non-null electromagnetic field, a null tetrad $(\m,\mc,\bl,\k)$ can always be chosen in
which the Maxwell tensor takes the simple form $F_{ab}=4\pho(m_{[a}\bar{m}_{b]}-k_{[a}l_{b]})$.
Then, the only non-trivial Ricci scalar is $\phoo\equiv\pho\phoc\neq0$. From the Bianchi identities
one obtains that the following spin coefficients vanish,
\be
 \kk=\s=\l=\n=\rr=\mm=\p=\t=0 \ .
\ee
Substituting these into the NP equations, it immediately follows  $R=0$. Thus, there are no such
solutions with $\Lambda\neq0$. Notice also that the Maxwell equations  have not been actually used,
but only the specific algebraic structure of the energy-momentum tensor.

\subsection*{Null fields}

In this case it is possible to generalize the approach used in \cite{McLTarTup75}.
Without loss of generality, we introduce a convenient null tetrad in which an arbitrary
null Maxwell field takes the form $F_{ab}=4\pht k_{[a}m_{b]}$, so that $\phtt\equiv\pht\phtc\neq 0$
is the only scalar. The Bianchi identities now imply
\beqn
 \kk & = & \s=\rr=0 \label{spin2} \ , \\
 D\phtt & = & -2(\e+\ec)\phtt \label{bia1} \ , \\
 \dec\phtt & = & (\tc-2\bc-2\a)\phtt \label{bia2} \ .
\eeqn
For our purposes, only three of the NP equations turn out to be necessary. Taking (\ref{spin2})
into account, these read
\beqn
 \de\a-\dec\b & = & \a\ac+\b\bc-2\a\b+\e(\mm-\mmc)+R/24 \ , \label{NP1} \\
 \de\t & = & (\t+\b-\ac)\t \label{NP2} \ , \\
 \dec\t & = & (\tc-\bc+\a)\t+R/12 \label{NP3} \ .
\eeqn
The Maxwell equations simplify to
\beqn
 D\pht & = & -2\e\pht \label{max1} \ , \\
 \de\pht & = & (\t-2\b)\pht \label{max2}\ .
\eeqn
From (\ref{bia2}) and (\ref{max2}), it follows
\be
 \dec\pht=-2\a\pht \ .
 \label{deltacphi1}
\ee
With (\ref{NP3})--(\ref{deltacphi1}) and (\ref{NP1}), the commutator
$(\dec\de-\de\dec)=(\mmc-\mm)D-(\ac-\b)\dec-(\bc-\a)\de$ applied on $\pht$ gives
\be
  \t\tc=-R/6 \ .
\ee
Since $\t\tc$ must thus be a constant, it follows $\de(\t\tc)=0$. Recalling (\ref{NP2}) and (\ref{NP3}),
this immediately leads to $0=(2\t\tc+R/12)\t=-R\,\t/4$ and therefore to $R=0$, which completes the
proof for null electromagnetic fields.

\addvspace{0.6cm}

The above results, which generalize the Theorems 32.16 and 32.17 of 
\cite{kramerbook}\footnote{Notice that Theorem 32.17 in \cite{kramerbook} by mistake includes not only null electromagnetic fields
but also generic pure radiation. This was pointed out in \cite{Wils89}, whereas the complete class of conformally
flat pure radiation metrics with $\Lambda=0$ was given in \cite{EdgLud97l}.}
to include a cosmological constant $\Lambda$, can be summarized  in the following
\begin{theorem}
 There are no conformally flat solutions of the Einstein--Maxwell equations (\ref{EM}) for $\Lambda\neq0$. When $\Lambda=0$, the only solutions are given by (\ref{BR}) and (\ref{plane}).
\end{theorem}

Let us  emphasize that the Maxwell equations (contrary to the case of non-null field) do play a key role
 in preventing the existence of conformally flat solutions with a \emph{null} electromagnetic field for $\Lambda\not=0$.
Indeed, there exist conformally flat metrics with an arbitrary $\Lambda$ and generic (i.e., non-electromagnetic)
pure radiation, as demonstrated in section \ref{sub_confflat}. A complete class of these spacetimes was
presented in \cite{OzsRobRoz85} (see also \cite{VanGunNar90} for the case $\Lambda>0$).
For $\Lambda=0$, such a general family splits into two invariant sub-classes, namely the Edgar--Ludwig metric
\cite{EdgLud97l} (still not representing electromagnetic fields) and the electromagnetic plane waves (\ref{plane}).

\addvspace{0.5cm}


\end{document}